\documentclass{aa}
\usepackage{graphicx}
\begin{document}
\title{Cold dust in a selected sample of nearby galaxies}
\subtitle{I. The interacting galaxy NGC\,4631}
\author{M. Dumke\inst{1,2}
\and    M. Krause\inst{1}
\and    R. Wielebinski\inst{1}
}
\institute{
    	Max-Planck-Institut f\"ur Radioastronomie,
	Auf dem H\"ugel 69, 53121 Bonn, Germany
\and    SMTO, Steward Observatory, The University of Arizona,
	933 N.~Cherry Avenue, Tucson, Arizona 85721, USA
	}
\offprints{M. Dumke,\\
\email{mdumke@mpifr-bonn.mpg.de}}
\date{Received 14 April 2003 / Accepted 17 October 2003}
\titlerunning{Cold dust in the interacting galaxy NGC\,4631}

\abstract{
We have observed the continuum emission of the interacting galaxy
NGC\,4631 at $\lambda\lambda$\,870\,$\mu$m and 1.23\,mm using the
Heinrich-Hertz-Telescope on Mt. Graham and the IRAM 30-m telescope
on Pico Veleta. We have obtained fully sampled maps which cover the
optical emission out to a radius of about $7'$ at both wavelengths.
For a detailed analysis, we carefully subtracted the
line contributions and synchrotron and free-free emission from the data,
which added up to 6\,\% at 1.23\,mm and 10\,\% at 0.87\,mm.
We combined the flux densities with FIR data to obtain dust spectra and
calculate dust temperatures, absorption cross sections, and masses.
Assuming a ``standard'' dust model, which consists of two populations of
big grains at moderate and warm temperatures, we obtained temperatures
of 18\,K and 50\,K for the both components.
However, such a model suffers from an excess of the radiation at
$\lambda\,1.23$\,mm, and the dust absorption cross section seems
to be enhanced by a factor 3 compared to previous results and theoretical
expectations. At large galactocentric radii, where the galaxy shows
disturbances as a result of gravitational interaction, this effect seems to
be even stronger.
Some possibilities to resolve these problems are discussed.
The data could be explained by a very cold dust component at a
temperature of 4\,--\,6\,K, an increased abundance of very small grains,
or a component of grains with unusual optical properties. We favour
the latter possibility, since the first two lead to inconsistencies.

\keywords{
galaxies: ISM -- radio continuum: galaxies -- galaxies: individual: NGC 4631
}
}

\maketitle

\section{Introduction}

The efficiency and time-scale of star-formation processes in galaxies
depend strongly on the amount, distribution, and composition of the
interstellar matter. One component of the ISM whose properties are
still not well known is interstellar dust.

Since most of the dust in galaxies is cold, with temperatures of
$< 20\,{\rm K}$, infrared observatories like IRAS
are almost blind for the gross amount of dust, and can only observe
the warm dust component. The major fraction of the interstellar dust
radiates mainly in the sub-mm range, at wavelengths $> 100\,\mu{\rm m}$.
This emission was hardly accessible in the past, due to the lack of good
sub-mm telescopes at sites with sufficiently good atmospheric conditions.
This situation
has improved in the last few years: Instruments like the bolometer
arrays installed at the 30-m telescope on Pico Veleta or - more
recently - SCUBA at the JCMT and the 19-channel array at the
Heinrich-Hertz-Telescope have produced important results for the
investigation of this cold dust component.
Also, the ISO satellite could routinely measure the peak of the cold dust
emission from grains with temperatures ranging down to 10~K
(e.g.\ Popescu et al.~\cite{popescu+02}).

However, dust properties in external galaxies are still not well
determined because of gaps in the spectral coverage. Another
difficulty in the analysis and interpretation of continuum data
is the possible contribution of molecular lines to the total flux
measured with the (broad-band) bolometer arrays, especially in
the usually used atmospheric windows around
$\lambda\lambda$\,870\,$\mu$m and
1.23\,mm which contain the strong CO(3--2) and (2--1) lines. This
situation currently improves as more and more objects are mapped
even in the higher CO transitions.

This current paper is the first in a series that reports on observations
of the thermal emission of cold dust in selected galaxies.
NGC\,4631 is a nearby ($D = 7.5\,{\rm Mpc}$, e.g.\ Golla \& Wielebinski
\cite{golla+rw94}) galaxy which is close to edge-on. While the position
angle of the large-scale (cm-) radio continuum emission is about
$86^{\circ}$, that of the inner disk is closer to $82^{\circ}$.
A value of $84^{\circ}$, which we assume throughout this paper, seems
to be most appropriate for the mm/sub-mm emission out to a radius of
a few arcminutes.
NGC\,4631 is embedded in a small gravitationally interacting group with
two neighboring galaxies. The dwarf elliptical galaxy NGC\,4627 is located
only $3'$ northwest of the nucleus of NGC\,4631, and another edge-on spiral,
NGC\,4656, is located about $30'$ to the southeast. This interaction has
created several prominent \ion{H}{i} bridges and spurs (Rand \cite{rand94}).
Since galaxy interactions play an important role in the evolution of
galaxy systems and the ISM content in galaxies, the investigation of
this object can provide useful information about many processes in the ISM.

NGC\,4631, which was classified as ``mild starburst'' by Golla \& Wielebinski
(\cite{golla+rw94}), may be in a late stage of its interaction, where
the central star formation (triggered by molecular inflow due to the
gravitational forces) has already ceased and an energetic outflow as well
as a huge radio halo have been formed
(Ekers \& Sancisi \cite{ekers+sancisi77}; Rand \cite{rand00}).

The observations of the dust component of NGC\,4631 started with the
IRAS satellite (Rice et al.\ \cite{rice+88}; Young et al.\ \cite{young+89}).
While the IRAS observations were sensitive mainly to the warm dust present
in this galaxy, the $\lambda$\,1.3\,mm map of Braine et
al.\ (\cite{braine+95}) proved the existence of a significant amount of
cold dust in the central area. The inner $2'\!\!.5$ were
also observed at $\lambda\,850\,\mu{\rm m}$ by Alton et
al.\ (\cite{alton+99}). Neininger \& Dumke (\cite{neininger+dumke99})
presented a more extended map at $\lambda$\,1.23\,mm (which was obtained
from the same data as shown in this paper) and detected
intergalactic cold dust which was pulled out of the disk by the gravitational
interaction of NGC\,4631 with its neighbours. Here we present a more
extended map at $\lambda\,870\,\mu{\rm m}$, covering the disk out to a
radius of $7'$.

Some basic parameters of NGC\,4631 are compiled in Table \ref{tab:basic}.

\begin{table}
\caption[]{Some basic parameters of NGC\,4631 as obtained from the literature.}
\label{tab:basic}
\begin{tabular}{lll}
\hline
Type & Sd & (de Vaucouleurs et al.\ \cite{rc3}) \\
\multicolumn{2}{l}{Position:} & (Young et al.\ \cite{young+95}) \\
RA[2000] & $12^{\rm h}42^{\rm m}07^{\rm s}\!\!.65$ & \\
Dec[2000] & $32^{\circ}32'27'\!\!.9$ & \\
Distance & 7.5\,Mpc & (Golla \& Wielebinski \cite{golla+rw94}) \\
Pos.\ Angle & $84^{\circ}$ & \\
Inclination & $86^{\circ}$ & \\
$M (\ion{H}{i})$ & $5 \times 10^9\,M_{\odot}$ & (Rand \cite{rand94}) \\
$M ({\rm H}_2)$ & $1.2 \times 10^9\,M_{\odot}$ & (Golla \& Wielebinski
\cite{golla+rw94}) \\
\multicolumn{2}{l}{IRAS flux densities:} & (Young et al.\ \cite{young+89}) \\
$12\,\mu{\rm m}$ & $  5.1 \pm  1.1$ & \\
$25\,\mu{\rm m}$ & $  8.8 \pm  1.8$ & \\
$60\,\mu{\rm m}$ & $ 90.7 \pm 18.2$ & \\
$100\,\mu{\rm m}$& $170.4 \pm 34.1$ & \\
\hline
\end{tabular}
\end{table}

\section{Observations and data reduction}
\label{section:obs}

We observed the galaxy NGC\,4631 at wavelengths
of $\lambda\lambda$\,870\,$\mu$m and 1.23\,mm, using sensitive
19-channel bolometer arrays developed by
E. Kreysa and collaborators at the Max-Planck-Institut f\"ur
Radioastronomie, Bonn.

\subsection{HHT observations}

The observations at $\lambda$\,870\,$\mu$m were carried out at the
Heinrich-Hertz-Telescope\footnote[1]{The HHT is operated by the
Submillimeter Telescope Observatory on behalf of Steward Observatory
and the MPI f\"ur Radioastronomie.}
(Baars et al.\ \cite{baars+99}),
located on Mt.\ Graham, Arizona, during three observing sessions between
March 2000 and January 2002, using a 19-channel bolometer array
installed as facility instrument.
The 19 channels of this bolometer are located in the centre and on the
sides of two concentric regular hexagons, with an apparent spacing
between two adjacent channels (beams) of $50''$. The central frequency
of the bolometer is about 345\,GHz (the highest sensitivity is reached
at 340\,GHz), and the instrument is sensitive mainly between 310 and
380\,GHz.

In order to calculate the atmospheric zenith opacity, we made
skydip observations every 45 to 90 minutes, depending on the atmospheric
stability. These yielded atmospheric opacities at our observing frequency
which varied between 0.3 and 0.9 for the several observing sessions.
For calibration purposes we have
also performed mapping and on-off measurements of various planets
(mainly Mars and Saturn) during the observations. These measurements
yielded a conversion factor from observed counts to mJy/beam of 
0.8 -- 1.1\,mJy\,beam$^{-1}$\,count$^{-1}$. This varying conversion factor
is due to (and partly corrects for) varying atmospheric condition at the
day of the observations and the uncertainties in the opacity calculation.
The beamwidth at this frequency is $\sim 23''$.

All maps were observed in the Az-El coordinate system, by scanning along
Azimuth and with data acquisition every 0.5\,s.
During the observations, the subreflector
was wobbled at 2\,Hz in azimuth, with a beam throw between $60''$ and
$200''$. If necessary, the starting point of each subscan was shifted by
a few arcseconds with respect to the preceding one, in order to ideally
place each individual coverage relative to the target source. This as
well as different map sizes (typically around $450'' \times 300''$) ensured
that each subscan covers the galaxy and a sufficient amount of blank sky
on either side, in order to facilitate baseline subtraction and a correct
restoration of the double-beam maps.

\subsection{Pico Veleta observations}

\begin{figure*}
\centering
  \includegraphics[angle=270,width=17cm]{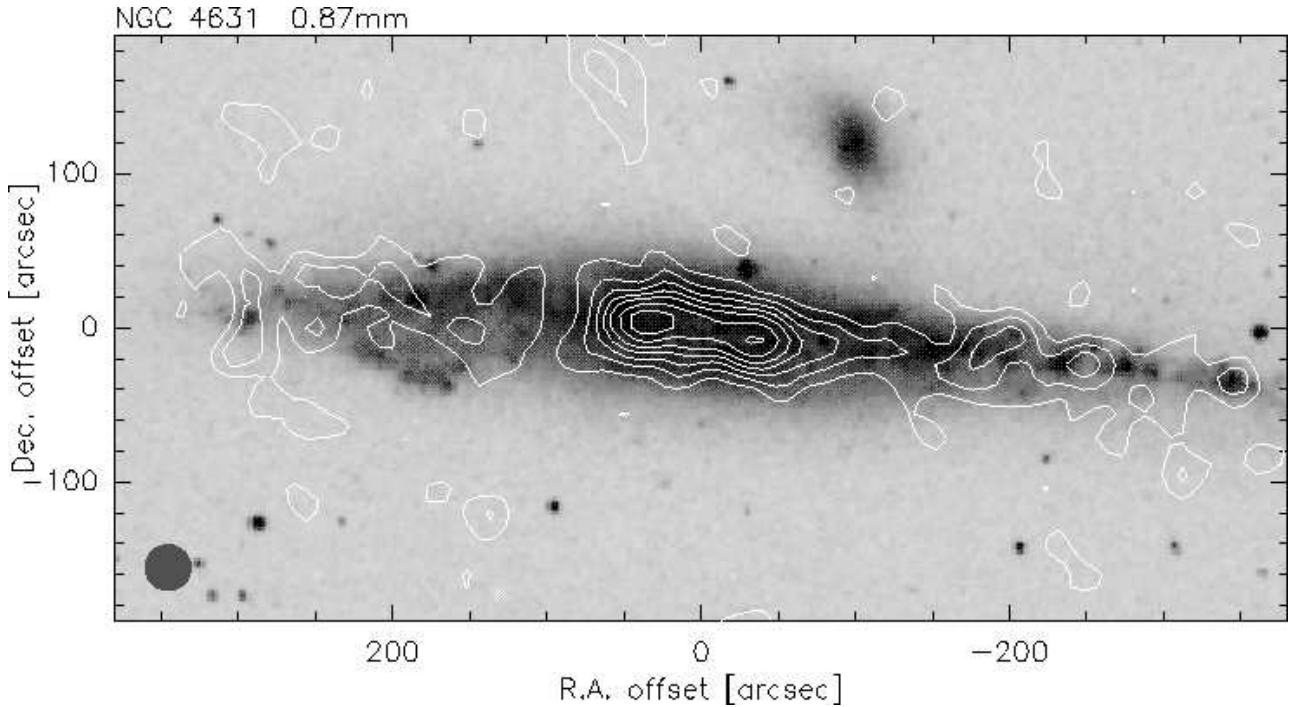}
  \caption{
   Continuum map of NGC\,4631 at $\lambda$\,0.87\,mm, overlaid on an
   optical image extracted from the Digitized Sky Survey. The map is
   smoothed to an angular resolution of $30''$; the beam size is
   indicated by the filled grey circle in the lower left corner.
   The rms noise in the map is about 20\,mJy/beam area, and
   contour levels are 40, 80, 120, 160, 200, 250, 300\,mJy/beam area.
   The dark spot in the optical (greyscale) image $3'$ northwest of the
   central area is the neighbouring galaxy NGC\,4627.
   }
  \label{fig:hht4631}
\end{figure*}

The observations at $\lambda$\,1.23\,mm were carried out in March 1997
at the IRAM 30-m telescope on Pico Veleta, Spain.
The central frequency of the used bolometer array is 243\,GHz,
the bandwidth about 70\,GHz.
The array layout is similar to the one at the
HHT, with individual channels separated by $20''$. The beamwidth
at the observing frequency is $11''$.

The sky opacity at the observing frequency was monitored with skydips and
varied between 0.1 and 0.3 during the five days of the observing run,
but was stable within $\pm 0.02$ for each individual day of the observations.
We mapped Mars every night to determine the absolute flux density scale;
from these measurements we obtained a conversion factor of
0.24 -- 0.28\,mJy\,beam$^{-1}$\,count$^{-1}$.

The observing procedures were similar to the HHT observations.
The maps were observed in the Az-El coordinate system, but with a lower
scanning velocity and smaller subscan separation in order to ensure full
sampling. The subreflector was wobbled at 2\,Hz with a beam throw of $45''$.

\subsection{Data reduction}

The data reduction for the HHT and the 30-m data was performed with the
NIC program of the GILDAS software package. In addition, we used the MOPSI
program to confirm (and adjust, if necessary) the zenith opacities
calculated by NIC from our skydip measurements, and also to estimate flux
densities for the observed planets in order to calibrate the data in
astronomical units.
After baseline subtraction and the elimination of spikes in each single
coverage, the atmospheric noise, which is highly correlated between the
individual channels (especially at $870\,\mu{\rm m}$), was subtracted.
The maps were gridded, restored (using the EKH algorithm), converted into
the RA-Dec system and finally combined (with an appropriate weighting)
to a single map for each of the two wavelengths.
The zero-levels of the resulting maps were checked and carefully adjusted.
Map features smaller than the telescope beam appearing in the final maps
were filtered out using a Fourier filter technique.
For further analysis, both maps were smoothed to a $FWHM$ of $24''$.
This was done in order to enable a direct comparison with the existing
CO data (see Sect.\ \ref{subsection:obs_contrib}) and other ISM
components. The rms noise level in the final maps is
20\,mJy/beam at $\lambda$\,0.87\,mm, and 3\,mJy/beam at $\lambda$\,1.23\,mm.

\section{Results}
\label{section:results}

\subsection{Total flux densities}

\begin{figure*}
\centering
  \includegraphics[angle=270,width=17cm]{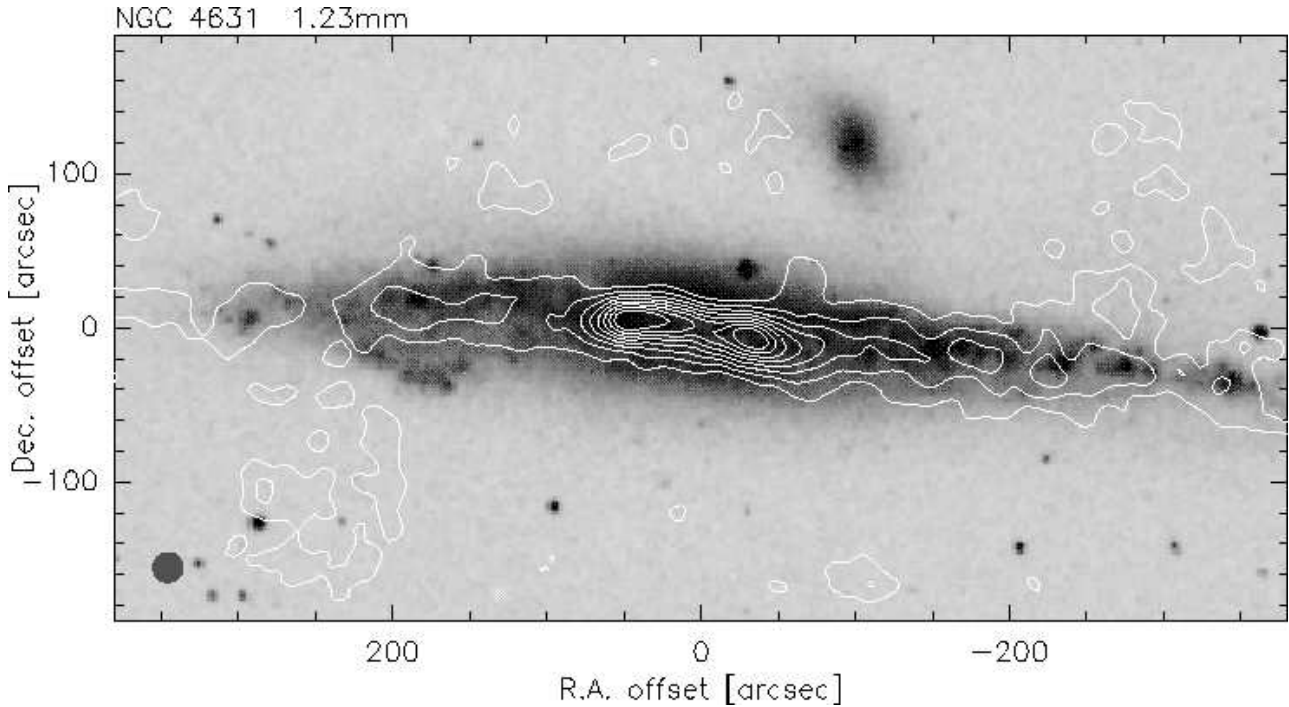}
  \caption{
   Continuum map of NGC\,4631 at $\lambda$\,1.23\,mm, overlaid on an
   optical image extracted from the Digitized Sky Survey. The map is
   smoothed to an angular resolution of $20''$; the beam size is
   indicated by the filled grey circle in the lower left corner.
   The rms noise in the map is about 3\,mJy/beam area, and
   contour levels are 10, 20, \ldots, 80\,mJy/beam area.
   }
  \label{fig:pv4631}
\end{figure*}
To measure the total flux densities, we used a ring integration
method which we applied to the disk of NGC\,4631. While the parameter
choice for this ring integration had some influence on the results, its
effect was much smaller than the uncertainties coming from the observations
and the data reduction process, including the calibration. Also the actual
rms noise in the final maps has a rather small effect. We estimate the
error of the flux densities to be 10\,\% at $\lambda$\,1.23\,mm, and
15\,\% at $\lambda$\,0.87\,mm.

The total flux density of NGC\,4631, including all disk emission, is
$S_{\rm 0.87mm} = 3.78 \pm 0.57\,{\rm Jy}$
for the observing wavelengths of 0.87\,mm.
At $\lambda$\,1.23\,mm, the total flux density is more difficult
to determine. The reason for this are the bridges and spurs which were
detected at this wavelength, as well as the fact that the edge of the
detected emission is more due to the edge of the mapped region rather than
due to the edge of the dust distribution. Limiting the integration of the
observed emission to the main disk of NGC\,4631, we estimate
$S_{\rm 1.23mm} = 2.18 \pm 0.22\,{\rm Jy}$. Note that we got these values
from integrating over the same area and using maps at the same angular
resolution ($24''$) for both wavelengths.

These values cannot be easily compared to previous flux density estimates,
since our maps are the first to cover (almost) all of the galaxy's emission.
Braine et al.\ (\cite{braine+95}) observed the central $4' \times 3'$ of
NGC\,4631 at a wavelength of 1.3\,mm and obtained a flux density of 0.64\,Jy.
When we restrict our integration to the same area, we measure $\sim 0.9$\,Jy.
This difference is only partly due to the slightly different wavelength.
Another reason for their lower value might be the lack of sensitivity for
the off-plane emission and therefore an imperfect baseline subtraction -- note
the negative areas north and south of the major axis and the overall
smaller intensity values in Fig.~1 of Braine et al.\ (\cite{braine+95}).

Bendo et al.\ (\cite{bendo+03}) used SCUBA archive data to estimate the
flux density at $\lambda\,850\,\mu{\rm m}$ and obtained
$1.89 \pm 0.19\,{\rm Jy}$ and $0.54 \pm 0.05\,{\rm Jy}$ for the central
$135"$ and $45"$, respectively. For the same areas, we get values of 1.8\,Jy
and 0.5\,Jy, which is in good agreement with those values.

\subsection{Morphology of the continuum maps}

The resulting maps of NGC\,4631 at $\lambda\lambda$\,0.87\,mm and 1.23\,mm
are shown in Figs.\ \ref{fig:hht4631} and \ref{fig:pv4631},
overlaid on an optical image extracted from the Digitized Sky Survey.

Similar to other ISM components, especially the CO molecular line emission,
both maps show a double-peaked appearance in the central area, which points
to a ring-like distribution with a diameter of about $1'$, or 2.2\,kpc at
the assumed distance of 7.5\,Mpc.

At larger radii, the emission is more patchy and follows the major axis of
the galaxy. The extent of the emission in the disk is difficult to
determine, since our map does not cover radii larger than $6' - 7'$ with a
sufficient signal-to-noise ratio.
At $\lambda$\,1.23\,mm we find a significant amount of emission at distances
of a few arcminutes away from the major axis, e.g.\ at
$(\Delta \alpha, \Delta \delta) = (250'', -150'')$ or $(0'', 120'')$,
and thus several kpc above the
plane. This emission is correlated with the \ion{H}{i} spurs found by Rand
(\cite{rand94}) and is discussed in detail by Neininger \& Dumke
(\cite{neininger+dumke99}). It is most likely interstellar material pulled
out of the plane by the gravitational interaction of NGC\,4631 with its
neighbouring galaxies. The total flux density of this extraplanar gas
-- as far as it is within our map edges -- is about 0.5\,Jy. This is
one quarter of the flux density within the plane of NGC\,4631, which
is 2.18\,mJy.

At $\lambda$\,0.87\,mm we are unable to detect this intergalactic cold
dust with sufficient significance. Only one area, located at
$(\Delta \alpha, \Delta \delta) = (60'', 170'')$, is clearly detected.
This feature
is located on the \ion{H}{i} spur pointing towards the northeast as seen by
e.g.\ Weliachew et al.\ (\cite{weliachew+78}). None of the other extraplanar
features of the 1.23\,mm map can be seen above the noise in the sub-mm map.
However, an integration of the map shows that a flux density of
$\sim 1\,{\rm Jy}$ originates in the off-plane features which can be
identified in the 1.23\,mm map, of which 400\,mJy are due to the emission
region described above, located northeast of the nucleus. The total flux
density confined to the plane of NGC\,4631 and located within our map is
$3.78 \pm 0.57\,{\rm Jy}$, so again the off-plane features contain about
25\,\% of the disk emission.

The radiation measured by the bolometers contain some contributions which
are not due to thermal dust emission. We calculate these non-dust
contributions in the next subsection, before we continue with the discussion
of the major axis distribution of important ISM components and the dust
spectrum.

\subsection{Origin of the observed emission}
\label{subsection:obs_contrib}

The broad band emission of the galaxies, measured with the bolometers,
consists of four main components: thermal dust emission,
free-free radiation from thermal electrons, synchrotron radiation from
relativistic electrons, and the CO as well as some weaker lines
which fall into the bandpass.
As we are mainly interested in the thermal dust emission alone,
we have to determine the contribution of the other processes and to subtract
them from the data.

Since CO mapping observations of NGC\,4631 exist, we can calculate
$B_{\rm CO}$, the contribution of the CO lines to the surface brightness
measured with the bolometers, through
\begin{equation}
B_{\rm CO} = {2k\nu^3c^{-3} \over \Delta\nu_{\rm bol}}\,\Omega_{\rm beam}
\,I_{\rm CO}\,,
\label{equ:B_CO1}
\end{equation}
where the line intensity $I_{\rm CO} = \int T_{\rm mb}\,dv$ denotes the
velocity-integrated main-beam brightness temperatures.
From the integrated CO line flux $F_{\rm CO}$, the contribution to the total
flux densities can be obtained via
\begin{equation}
S_{\rm CO} = {\nu c^{-1} \over \Delta\nu_{\rm bol}}\,F_{\rm CO}\,.
\end{equation}
Here the equivalent bandwidth of the bolometer can be calculated from
the instrumental bandpass (Kreysa, priv.\ comm.) and is
$\Delta\nu_{\rm bol} \sim 50\,{\rm GHz}$ for the CO(3--2) line and
$\Delta\nu_{\rm bol} \sim 70\,{\rm GHz}$ for the CO(2--1) line.

The contribution of the CO(3--2) to the 870\,$\mu$m flux can be
taken from data published by Dumke et al.\ (\cite{dumke+01}), and
the CO(2--1) contribution to the 1.23\,mm flux can be estimated from
the data published by Golla \& Wielebinski (\cite{golla+rw94}).
We should note that at both wavelengths the CO emission was observed
with the same telescope as the corresponding continuum emission, thus
eliminating possible uncertainties due to beam characteristics.
The total line contribution, including also $^{13}$CO and other
lines, may be somewhat higher. Based on typical line ratios in
external galaxies and the sensitivity variation of the bolometer over
the bandpass, we estimate that other lines add up to about 5\,\% of the
flux calculated from the $^{12}$CO(3--2) line, and about 10\,\% of the
flux from the $^{12}$CO(2--1) line.
Taking this into account and entering all quantities in
Eq.\ (\ref{equ:B_CO1}), it simplifies to
\begin{equation}
B_{\rm line}^{\rm 0.87mm}\,{\rm [mJy]}
 = 1.36\,I_{\rm CO(3-2)}\,{\rm [K\,km\,s}^{-1}{\rm ]}
\end{equation}
and
\begin{equation}
B_{\rm line}^{\rm 1.23mm}\,{\rm [mJy]}
 = 0.303\,I_{\rm CO(2-1)}\,{\rm [K\,km\,s}^{-1}{\rm ]}
\end{equation}
for the $\lambda\lambda\,870\,\mu{\rm m}$ and 1.23\,mm emission,
respectively, and a $24''$ beam.
To correct for the line contribution in the continuum maps, we scaled the
CO intensity maps accordingly and subtracted these from the bolometer maps.
For a comparison of the dust distribution at both wavelengths with other
ISM components (see next subsection) we smoothed all maps to a final
resolution of $24''$ HPBW.

\begin{figure*}
  \includegraphics*[width=12cm]{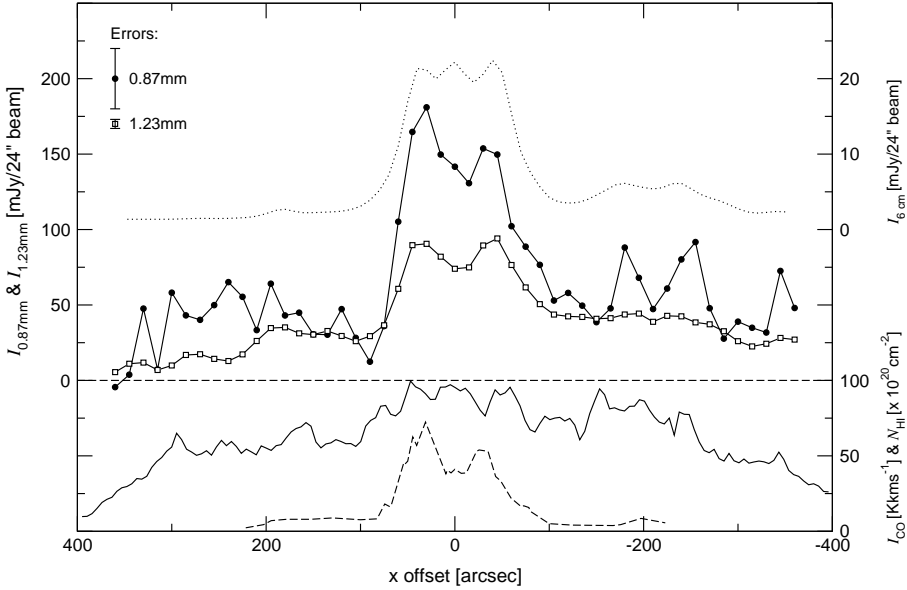}
    \caption{Distribution of various ISM components along the major axis
    of NGC\,4631. The thermal dust emission at $870\,\mu{\rm m}$ and 1.23\,mm,
    as obtained from the maps in Figs.\ \ref{fig:hht4631} and \ref{fig:pv4631}
    and corrected for line, synchrotron and free-free contributions, are shown
    as filled circles and white squares. The errors of both data sets are
    indicated in the upper left corner, while the long-dashed line gives their
    zero level.
    Also shown for comparison are the \ion{H}{i} profile (solid line),
    the CO(1--0) distribution (dashed line),
    and the $\lambda$\,6\,cm radio continuum
    distribution (dotted line). The scale for these
    three data sets is shown at the y-axis on the right.
    }
    \label{fig:dist1}
\end{figure*}

For the line contributions to the total flux denity, we get
$S_{\rm line}^{\rm 0.87mm} = 330 \pm 30\,{\rm mJy}$ and
$S_{\rm line}^{\rm 1.23mm} = 90 \pm 10\,{\rm mJy}$.
The fraction of this contribution to the observed flux density is thus
9\,\% and 4\,\% for the two wavelengths of 0.87\,mm and 1.23\,mm,
respectively.

The fraction of the free-free and the synchrotron emission
can be estimated from radio continuum data at lower frequencies.
Niklas et al.\ (\cite{niklas+95}, \cite{niklas+97}) separated these
two components in the radio spectra of a large sample of galaxies.
At a frequency of 10\,GHz they found a thermal fraction of the emission
of ${\rm f}_{\rm th} = 0.13$ and a nonthermal spectral index of
$\alpha_{\rm nth} = 0.78$.
With these values, we estimate a contribution of the free-free and
synchrotron emission at both wavelengths of $40 \pm 10\,{\rm mJy}$
(0.87\,mm) and $50 \pm 10\,{\rm mJy}$ (1.23\,mm), which is of the order
of 1 -- 2\,\% of the total emission.
Although the thermal fraction of the cm-emission might be underestimated
because of galactic wind and cosmic ray propagation effects on the
integrated radio spectrum (Werner \cite{werner88}), the contribution
of free-free and synchrotron radiation is still less than 80\,mJy at
both wavelengths even in the unlikely case that more than half of the
emission in the cm-range is free-free emission.
Therefore it is much smaller than
the line contribution, especially at $\lambda$\,0.87\,mm. Thus any
uncertainties in the assumed thermal (free-free) fraction of the
cm-emission and the non-thermal spectral index are negligible for the further
data analysis, considering the total uncertainty of the flux density values
of 10 -- 15\,\%.

The contributions of the line emission and the non-dust continuum radiation
to the total flux densities at 0.87\,mm and 1.23\,mm are listed in
Table 2.
At this point we should note that the main uncertainties in the results are
due to the rms noise in and the absolute calibration of the continuum maps.
The subtraction (or non-subtraction) of the non-dust contribution does not
change our results qualitatively, and the uncertainties in these contributions
are negligible.

\begin{table}
\caption{Flux densities of NGC\,4631 at $\lambda\lambda\,0.87$\,mm and
1.23\,mm.
The non-dust contributions at both wavelengths are calculated as
described in Sect.\ \ref{subsection:obs_contrib}.}
\label{tab:fluxes}
\tabcolsep3pt
\begin{tabular}{ccccc}
\hline
$\lambda$ & $S_{\rm obs}$ [Jy] & $S_{\rm line}$ [mJy]
	& $S_{\rm ff+sync}$ [mJy] & $S_{\rm dust}$ [Jy] \\
\hline
0.87\,mm & $3.78 \pm 0.57$ & $330 \pm 30$  & $40 \pm 10$ & $3.41 \pm 0.58$ \\
1.23\,mm & $2.18 \pm 0.22$ &  $90 \pm 10$  & $50 \pm 10$ & $2.04 \pm 0.23$ \\
\hline
\end{tabular}
\end{table}

\subsection{ISM distribution along the major axis}
\label{subsection:majoraxis}

We determined the distribution of various components of the ISM along the
major axis of NGC\,4631, including the current results from
$\lambda\lambda$\,0.87 and 1.23\,mm (after subtraction of the non-dust
contributions), the \ion{H}{i} (Rand \cite{rand94}),
the CO(1--0) (Golla \& Wielebinski \cite{golla+rw94}), and the
$\lambda$\,6\,cm radio continuum emission (Krause et al., in prep.).
The result is shown in Fig.\ \ref{fig:dist1}, with an angular resolution
of $24''$ for all data sets.
Here we averaged the data over three pixels (pixel size is $8''$)
perpendicular to the major axis
for each data point in order to increase the signal-to-noise ratio, and to
account for emission which is close to, but not exactly on the major axis.
The $\lambda$\,6\,cm data are included as a tracer for the FIR emission,
since at this wavelength we see mainly non-thermal radiation, for which
the radio-FIR correlation is strongest.

Similar to other edge-on galaxies, the dust emission resembles the
distribution of molecular gas in the inner part of the galaxy. However,
at radii where no molecular gas can be detected ($r > 4'$), there is
still a significant amount of thermal dust emission, which in this
case follows the distribution of the \ion{H}{i}.

By comparing the distributions for the various wavelengths of radio continuum
emission, we note some significant differences between the data sets.
The central maximum is only visible in the $\lambda$\,6\,cm data, but not at
mm/sub-mm wavelengths.
Since the existence of an active nucleus, capable to produce excess
cm emission, is rather unlikely in NGC\,4631, this points to a steeper
FIR-to-mm spectrum and thus to higher dust temperatures in the very centre.
Also the intensity of the ``dust ring'' is different at the eastern and western
maximum, suggesting different dust properties in these two maxima.
All distributions show a steep decrease in intensity at the
eastern edge of the molecular ring, at a radius of $40'' - 70''$,
while the intensity decrease at the western edge is much shallower. 

At the molecular ring itself, at a radius of about $30''$, all continuum
distributions show local maxima. Such a correlation between molecular gas
and dust is expected because many molecules are formed on the surface of dust
grains, and the dust shields the gas from starlight and prevents the
molecules from being photodissociated. However, the \ion{H}{i} distribution
shows local minima at these positions.
Here the interstellar hydrogen was probably transformed into molecules because
of a local increase of ambient density. This is a direct consequence of gas
flows due to a non-axisymmetric potential, which results from the gravitational
interaction with the neighbouring galaxies NGC\,4656 and NGC\,4627.

While the CO intensities have dropped to relatively low values beyond the
molecular ring, the \ion{H}{i} emission stays at significant levels.
Also the tracers of interstellar dust ($\lambda\lambda$\,1.23\,mm, 0.87\,mm,
and 6\,cm, i.e. FIR) are still detected at larger radii.
However, different correlations between the various ISM components can be
observed. In the western disk, $3' - 4'$ away from the centre, the
\ion{H}{i} distribution shows a plateau-like feature. Here also the
0.87\,mm and 6\,cm data show maxima in their distribution, while we
see only a small increase in the 1.23\,mm data.
In the eastern disk, on the other hand, the 1.23\,mm and 6\,cm distributions
drop beyond a radius of $3'$, while the 0.87\,mm emission stays at a
higher level, similar to the \ion{H}{i}. These significant differences
in the distribution of the various dust tracers point to changing
dust properties along the major axis of NGC\,4631. This will be discussed
in more detail in Sect.\ \ref{subsection:cross}.

\section{Dust properties in NGC\,4631}

\subsection{The observed dust spectrum}

Table \ref{tab:fluxes}
lists the observed flux densities for both wavelengths. In addition, it
gives the non-dust contributions to the flux densities and the resulting
values which are due to the thermal emission of dust alone.
As discussed in Sect.\ \ref{subsection:obs_contrib},
the line contribution of the measured continuum flux densities is
9\,\% at $\lambda\,0.87$\,mm, and 4\,\% at $\lambda\,1.23$\,mm.
The latter value is significantly below this value measured in other
galaxies,
either normal (e.g.\ NGC\,5907, Dumke et al.\ \cite{dumke+97}),
active (e.g.\ NGC\,3079, Braine et al.\ \cite{braine+97}), or
starbursting (e.g.\ M\,82, Thuma et al.\ \cite{thuma+00}).
There the line contribution to the measured continuum flux ranges from
8\,\% in NGC\,5907 to 35\,\% in the centre of NGC\,3079, and it seems
to be correlated with the activity status of the galaxy. This would point
to NGC\,4631 being a rather inactive galaxy. However, a small fraction of
line emission can also point to a higher value for the dust emission per
gas mass, i.e.\ to a higher absorption cross section (see further below).

The two measured flux density values at $\lambda\lambda$\,0.87 and 1.23\,mm,
together with the IRAS data (Young et al.\ \cite{young+89}), still leave
a large gap in the spectral coverage between the FIR and the sub-mm.
In order to fill this gap, we take the ISO and SCUBA data published by Bendo
et al.\ (\cite{bendo+02}, \cite{bendo+03}) and estimate flux density values
for wavelengths of $180\,\mu{\rm m}$ and $450\,\mu{\rm m}$.

Concerning the SCUBA data, these authors give a flux density value only for the
inner $135''$ of the galaxy. In order to estimate
the {\it total} flux density for NGC\,4631, we have to scale this by a factor
$S_{\rm tot}/S_{135''}$, describing the ratio of the total flux density to
the flux density in the central $135''$ of the galaxy. From our maps, we
measured this ratio to $2.27$ and $2.34$ at a wavelength of 0.87\,mm and
1.23\,mm, respectively (after subtraction of the non-dust contributions,
which are negligible at $\lambda\,450\,\mu{\rm m}$).
In the FIR, the ratio of IRAS flux densities (corresponding to the total emission)
to the ISO values (corresponding to the inner $135''$)
is about 1.6 at $\lambda\,60\,\mu{\rm m}$ and 1.5 at
$\lambda\,100\,\mu{\rm m}$, where the latter value is less certain due to the
wider PSF of ISOPHOT at this wavelength. Considering that
the warmer dust (i.e.\ at shorter wavelengths) is probably more concentrated
to the inner disk than the colder dust, we assume a factor of
$S_{\rm tot}/S_{135''} = 2.0$ at
$\lambda\,450\,\mu{\rm m}$. With this value we get a total flux density of
$S_{450\mu{\rm m}} = 36 \pm 9\,{\rm Jy}$.

In order to estimate the total flux density at $\lambda\,180\,\mu{\rm m}$, we
take into account that the ISOPHOT C200 detector array covers only the central
$180'' \times 180''$ of the galaxy, and therefore also misses a significant
fraction of the disk emission. Again we can estimate a scaling factor from our
maps, and find $S_{\rm tot}/S_{180''} \sim 1.8$ for $\lambda\,180\,\mu{\rm m}$.
Interestingly, we find at $\lambda\lambda$\,0.87\,mm and 1.23\,mm that the flux
density within the central $180''$ is not much larger than that within the
central $135''$. This can be understood by the fact that the smaller $135''$
area covers already the central ``dust ring'', as described in the previous
section. This result differs from the values given by
Bendo et al.\ (\cite{bendo+03}), who found a much
larger difference between the two flux density values for the inner $135''$
and the inner $180''$. This inconsistency can probably
be explained by the fact that these authors used a SCUBA sub-mm map for
a deconvolution analysis, although this map covers less then $3'$ of the inner
disk, and is therefore not well suited for this purpose.

In addition, the spatial resolution of ISO at this wavelength is not much
smaller than the detector array.
With the knowledge of the PSF of this detector one can determine which fraction
of the total emission of a source would be detected by the C200.
For a somewhat extended source like NGC\,4631 we estimate a value of 82\,\%.
Taking these two corrections into account, we get a total flux density
at this wavelength of $S_{180\mu{\rm m}} = 205 \pm 35\,{\rm Jy}$.

\begin{figure}
  \resizebox{\hsize}{!}{\includegraphics{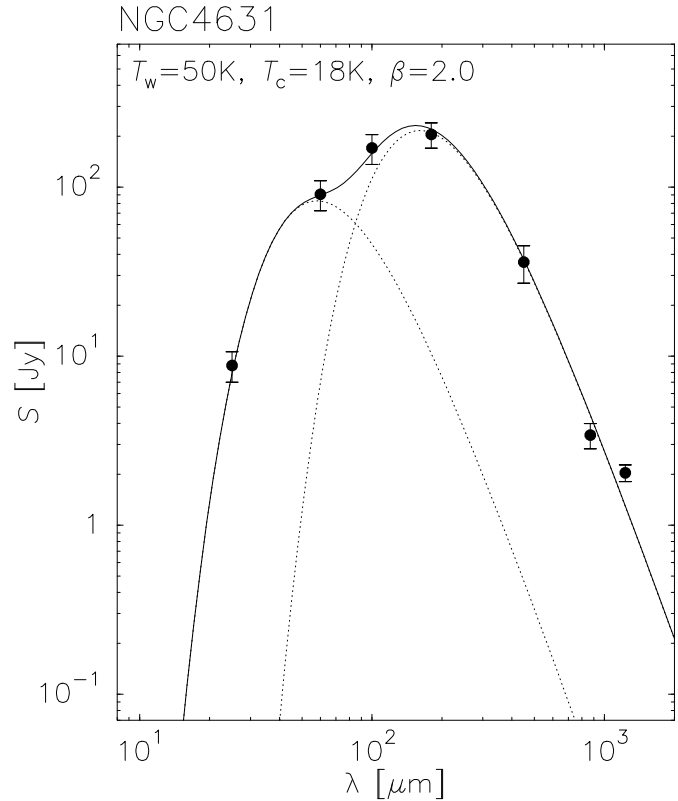}}
  \caption{The FIR-to-mm spectrum of NGC\,4631. The FIR data points up to
  $\lambda\,100\,\mu{\rm m}$ are taken from Young et al.\ (\cite{young+89}),
  those at $180\,\mu{\rm m}$ and $450\,\mu{\rm m}$ are estimated from the
  values in Bendo et al.\ (\cite{bendo+03}), and the 0.87\,mm and 1.23\,mm
  flux densities are from this paper and corrected for line contributions
  and synchrotron and free-free emission. The solid line shows a fit for
  a ``standard'' dust model (two modified Planck curves with $\beta = 2$).
  }
  \label{fig:spectrum1}
\end{figure}

\subsection{Dust temperatures and composition}
\label{subsection:tempandcomp}

The interstellar dust in galaxies is a mixture of many components at
several temperatures. Although the exact composition and temperature
distribution depends on the local interstellar radiation field and local
processes important for dust grain processing, the total FIR to
mm-spectrum of a galaxy can usually be fitted well by 1 -- 3 components
with a modified black-body spectrum of the form
\begin{equation}
S_{\lambda} \propto \sigma_{\lambda}^{\rm H}\,B_{\lambda}(T_{\rm d})\,,
\end{equation}
with the dust absorption cross section per hydrogen atom
$\sigma_{\lambda}^{\rm H} \propto \lambda^{-\beta}$. The value of $\beta$
depends on the dust composition and is still a matter of debate;
theoretical arguments suggest values between 1 and 2.
Realistic dust models, which are able to explain the observed dust
emission and extinction, suggest $\beta \sim 2$ for big grains which radiate
in thermal equilibrium with the ambient temperature
(Andriesse \cite{andriesse74}; Draine \& Lee \cite{draine+lee84}).
In fact the FIR-to-mm spectra of many external galaxies have been
successfully fitted with one or two components of dust with $\beta = 2$
(e.g.\  Chini et al.\ \cite{chini+95};
Neininger et al.\ \cite{neininger+96};
Dumke et al.\ \cite{dumke+97};
Braine et al.\ \cite{braine+97};
Alton et al.\ \cite{alton+98};
Stevens \& Gear \cite{stevens+gear00}).

We fitted the FIR-to-mm spectrum of NGC\,4631 with a dust model consisting
of two components, both with $\beta = 2$.
We used data points from Young et al.\ (\cite{young+89}) in the FIR
(25 -- 100\,$\mu$m), those estimated above at $180\,\mu{\rm m}$ and
$450\,\mu{\rm m}$ from Bendo et al.\ (\cite{bendo+03}), and our new values
at 0.87\,mm and 1.23\,mm. The result is shown in Fig.\ \ref{fig:spectrum1},
where the solid curve represents a two-component modified Planck spectrum
with temperatures of 50\,K and 18\,K. This temperatures agree well with
previous results (see references above), where the cold dust component was
usually found to have a temperature of 15 -- 20\,K.

We can also estimate dust temperatures for the inner and outer disk separately
when we use the flux density values at $\lambda\lambda\,60\,\mu$m and $100\,\mu$m
from Bendo et al.\ (\cite{bendo+03}) for
the inner $135''$, and restrict the integration in our maps to the same area.
A fit to the data leads to somewhat higher dust temperatures of 51\,K and
20\,K for the warm and cold component, respectively. After subtracting the
flux density values for the inner disk from the total flux densities, we also
estimate dust temperatures for the outer disk and find 49\,K for the warm
component and 16\,K for the cold component. Please note, however, that
this two-component model is still a model, and the main conclusion from these
fits is that the dust temperature decreases from the centre of NGC\,4631 to the
outer disk.

As it can be seen in Fig.\ \ref{fig:spectrum1}, this standard model
seems to be not sufficient to explain the measured flux densities at
both 0.87 and 1.23\,mm. Instead we detect an excess in the mm-range
(or, alternatively, a lack of emission in the sub-mm):
The ratio of the two dust-only flux densities (i.e.\ after
subtraction of line and other contributions) is
$S_{\rm dust}({\rm 0.87\,mm})/S_{\rm dust}({\rm 1.23\,mm}) = 1.67 \pm 0.35$,
while we would expect a ratio of 4 in case of a modified Planck spectrum
with a dust absorption coefficient $\propto \lambda^{-2}$. This
excess of dust emission at 1.23\,mm leads to a flattening of the
sub-mm/mm spectrum of NGC\,4631, which is difficult to explain with the thermal
radiation of big dust grains at moderately warm temperatures alone.
However, for practical reasons, we continue here with the analysis of the
data by assuming that the above model is reasonable, and discuss this
mm excess further below.

\subsection{Absorption cross sections}
\label{subsection:cross}

An averaged value of the dust absorption cross section in NGC\,4631 can be
calculated from the measured flux density and the hydrogen mass through
(e.g.\ Hildebrand \cite{hildebrand83})
\begin{equation}
\sigma_{\lambda}^{\rm H} =
{S_{\lambda}\,m_{\rm H}\,D^2
\over B_{\lambda}(T_{\rm d})\,M_{\rm H}}\,.
\label{equ:SigmaofMH}
\end{equation}
Here $m_{\rm H}$ is the mass of a hydrogen atom, $D$ the distance to the
galaxy, and $M_{\rm H} = M_{\ion{H}{i}} + M_{\rm H_2}$ the hydrogen mass.
From observations of the atomic hydrogen in NGC\,4631, Rand (\cite{rand94})
finds a \ion{H}{i} mass of the disk (excluding the Helium content) of
$M_{\ion{H}{i}} = 5 \times 10^9\,M_{\odot}$. We don't take into account here
the mass of the various \ion{H}{i} spurs detected by Rand (\cite{rand94}),
which would add another $2.2 \times 10^9\,M_{\odot}$.
The molecular mass found by Golla \& Wielebinski (\cite{golla+rw94}) is
$M_{\rm H_2} = 1.2 \times 10^9\,M_{\odot}$, based on the galactic
CO-H$_2$ conversion factor of
$X = 2.3 \times 10^{20}\,{\rm cm}^{-2}/({\rm K\,km\,s}^{-1})$
(Strong et al.\ \cite{strong+88}). Although we know that $X$ may be variable
and not necessarily valid for NGC\,4631, the fraction of the molecular mass
is rather small in this galaxy, thus we use
$M_{\rm H} = 6.2 \times 10^9\,M_{\odot}$.
When we take this value for the hydrogen mass, assume a dust temperature of
18\,K as determined above (note that at our observing wavelengths around
1\,mm more than 99\,\% of the dust emission comes from the cold
component; see Fig.\ \ref{fig:spectrum1}), and use
Eq.\ (\ref{equ:SigmaofMH}) to calculate the absorption cross section for
the whole galaxy, we get values of
$\sigma^{\rm H}_{\rm 0.87mm} = 6.2 \pm 1.0 \times 10^{-27}\,{\rm cm}^2$
and $\sigma^{\rm H}_{\rm 1.23mm} = 6.4 \pm 0.8 \times 10^{-27}\,{\rm cm}^2$
for both wavelengths.

In order to interpret these values, we use a formulation following
Mezger et al.\ (\cite{mezger+90}),
\begin{equation}
\sigma^{\rm H}_{\lambda} = C\,\lambda_{\rm mm}^{-2}\,b\,Z/Z_{\odot}\,.
\label{equ:sigmaZ}
\end{equation}
$C = 7 \cdot 10^{-27}\,{\rm cm}^2$ is the dust absorption cross section at
$\lambda$\,1\,mm following from the theoretical curves of Draine \& Lee
(\cite{draine+lee84}), $\lambda_{\rm mm}$ the wavelength in mm, $Z$ the
metallicity and $b$ an empirically determined factor which accounts for the
differences between Drain \& Lee's grain mixtures and real grains.
$b = 1$ applies to dust in the diffuse interstellar \ion{H}{i} gas, and since
most gas in NGC\,4631 is in atomic form, we will also use $b = 1$ here.
For solar metallicity we would thus expect values of
9.2 and 4.6\,$\times 10^{-27}\,{\rm cm}^{2}$ for 0.87 and 1.23\,mm,
respectively. At the latter wavelength, such a value has been measured for
several nearby galaxies (e.g.\ Neininger et al.\ \cite{neininger+96};
Dumke et al.\ \cite{dumke+97}).
Kr\"ugel \& Chini (\cite{kruegel+chini94}) give a ``standard'' value which
corresponds to $\sigma_{\rm 1.3mm}^{\rm H} = 5.0 \times 10^{-27}\,{\rm cm}^2$
based on typical abundances of silicate and graphite in the ISM,
which is in very good agreement with the values above and has been used
successfully for fitting IR spectra over the whole spectrum from 1 to
1300\,$\mu$m.

To compare our results with these values, we have to take the metallicity into
account: Otte et al.\ (\cite{otte+02}) measured the Nitrogen abundance in
NGC\,4631 and found an average metallicity of $Z/Z_{\odot} \sim 0.5$. This is
also in agreement with the results from Vila-Costas \& Edmunds
(\cite{vilacostas+edmunds92}) who determined metallicities and their gradients
in Sd galaxies. With $Z = 0.5\,Z_{\odot}$, the observed absorption cross
section $\sigma$ is a factor 1.5 higher than the predicted value for
$\lambda$\,0.87\,mm, and even a factor 3 higher than the predicted value
for $\lambda$\,1.23\,mm.

The dust absorption cross section can also be expressed by
(Hildebrand \cite{hildebrand83})
\begin{equation}
\sigma^{\rm H}_{\lambda}
= \kappa_{\lambda}\,m_{\rm H}\,{M_{\rm d} \over M_{\rm g}}
\label{equ:sigmakappa}
\end{equation}
with the dust absorption coefficient $\kappa_{\lambda}$ (which involves
grain properties like size, density, and emissivity) and the dust-to-gas
mass ratio $M_{\rm d}/M_{\rm g}$. The high values for
$\sigma$ can thus be explained either with a high dust-to-gas ratio (which
may be in contradiction to the low metallicity) or an increased absorption
coefficient in the sub-mm/mm regime, and therefore unusual optical
properties of the dust grains in NGC\,4631.

From Eq. (\ref{equ:SigmaofMH}) it is obvious that
these high values for $\sigma$ are due to rather high flux densities.
But when comparing the numbers for both wavelengths, the high value at
$\lambda$\,1.23\,mm can also be directly related to
the unusual ratio between the two flux densities at our observing
wavelengths. On the other hand, as shown in Fig.\ \ref{fig:dist1}, we also
see that the intensity ratio between these two wavelengths varies strongly
along the galaxy's major axis.
Therefore we also calculate the absorption cross section {\it locally} for
selected areas along the major axis of NGC\,4631, in order to trace local
variations of the unusual dust properties found in the total spectrum.

The flux density per beam emitted by a cloud of gas and dust is
(see Mezger et al.\ \cite{mezger+90})
\begin{equation}
S_{\lambda} = \Omega_{\rm beam} B_{\lambda}(T_{\rm d})
(1 - {\rm e}^{-\tau_{\lambda}})
\label{equ:Sigma_Mezger}
\end{equation}
with $\tau_{\lambda} = \sigma^{\rm H}_{\lambda} N_{\rm H}$, where
$\sigma^{\rm H}_{\lambda}$ is the dust absorption cross section per hydrogen
atom and $N_{\rm H}$ the beam-averaged hydrogen column density.
For $\tau_{\lambda} \ll 1$ we can transform Eq.\ (\ref{equ:Sigma_Mezger})
and express the cross section by
\begin{equation}
\sigma^{\rm H}_{\lambda} =
 {\lambda^2 \over \Omega_{\rm beam}\,2kT_{\rm d}}\,
 {({\rm e}^x - 1) \over x}\,{S_{\lambda} \over N_{\rm H}}
\label{equ:Sigma_optduenn}
\end{equation}
with $x = hc/\lambda k T_{\rm d}$ and
$N_{\rm H} = N(\ion{H}{i}) + 2\,N({\rm H}_2)$.

\begin{figure}
\centering
  \resizebox{\hsize}{!}{\includegraphics[clip=]{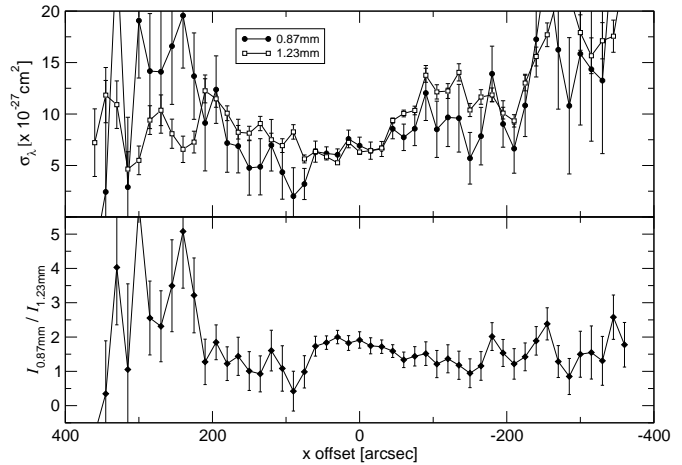}}
  \caption{Distribution of the intensity ratio of the two observing
  wavelengths (lower panel) and the absorption cross sections for a standard
  dust model (upper panel) along the major axis of NGC4631.
  }
  \label{fig:ratsig}
\end{figure}
We can use Eq.\ (\ref{equ:Sigma_optduenn}) to calculate the absorption
cross section in various parts of NGC\,4631. For this calculation we assume
the temperature of the cold dust to decrease linearly from 20\,K in the centre
to 16\,K in the outer parts, as found in the previous subsection.
The result is shown in the upper panel of Fig.\ \ref{fig:ratsig}, which
also shows (in the lower panel) the variation of the intensity ratio
$I_{\rm 0.87mm}/I_{\rm 1.23mm}$ along the major axis.

Before we discuss these data, we average the results in various parts of
the major axis in order to increase the S/N: in the centre (strong CO,
$|x| < 1'\!\!.5$), in the inner disk (weak CO, $1'\!\!.5 \le |x| < 4'$), and
in the outer disk (no detected CO, $|x| \ge 4'$). The obtained numbers
for the dust intensity ratios and absorption cross sections are given
in Table \ref{tab:sigma}.

\begin{table}
\caption[]{Dust emission intensity ratios and absorption cross section in
various regions along the major axis of NGC\,4631.}
\label{tab:sigma}
\begin{tabular}{llll}
\hline
 Position & $I_{\rm 0.87mm}/I_{\rm 1.23mm}$
 & $\sigma^{\rm H}_{\rm 0.87mm}$ & $\sigma^{\rm H}_{\rm 1.23mm}$ \\
 & & \multicolumn{2}{c}{$[\times 10^{-27}\,{\rm cm}^{2}]$} \\
\hline
$x \ge 4'$ (east)          & $2.8 \pm 0.6$ & $15 \pm 4$ & $ 8 \pm 2$ \\
$4' > x \ge 1'\!\!.5$      & $1.3 \pm 0.2$ & $ 7 \pm 2$ & $ 9 \pm 1$ \\
$1'\!\!.5 > x > -1'\!\!.5$ & $1.7 \pm 0.1$ & $ 7 \pm 1$ & $ 7 \pm 1$ \\
$-1'\!\!.5 \ge x > -4'$    & $1.4 \pm 0.1$ & $ 9 \pm 2$ & $12 \pm 1$ \\
$x \le -4'$ (west)         & $1.7 \pm 0.2$ & $19 \pm 3$ & $18 \pm 2$ \\
\hline
\end{tabular}
\end{table}

From these numbers, we can roughly distinguish three different situations.
In the centre and inner disk, the absorption cross section at
$\lambda\,0.87$\,mm is only slightly higher than expected, whereas it is
much higher at $\lambda\,1.23$\,mm. Also the intensity ratio is very low,
especially in the areas of weak CO emission, where $\sigma_{\rm 1.23mm}$
is even higher than $\sigma_{\rm 0.87mm}$.
In the outer parts however, where the optical image of NGC\,4631 shows strong
disturbances, the situation has changed. At the eastern edge of the disk,
also $\sigma^{\ion{H}{i}}_{\rm 0.87mm}$ is strongly enhanced, while the
intensity ratio between both wavelengths is closer to the expected values.
At the western edge, on the other hand, the intensity ratio is similar to the
central area, but at both wavelengths the absorption cross section is
increased by another factor 2 compared to the central area and inner disk.

To summarize the results of this subsection, we find that the dust
absorption cross section in the sub-mm/mm range is higher than
what is predicted, and this result is especially significant in the
outer disk ($|x| > 4'$) of NGC\,4631, where no molecular gas is
detected. These areas are also strongly affected by the gravitational
interaction of NGC\,4631 with its neighbours. The optical image reveals
strong disturbances of the disk, and also the footpoints of two \ion{H}{i}
spurs detected by Rand (\cite{rand94}) are located in these areas.
A significant amount of molecular gas in these regions without
corresponding CO emission could be responsible for the higher values
of the absorption cross section at $\lambda\,0.87$\,mm, since this would
result in an underestimate of $N_{\rm H}$ and therefore a too high value
of $\sigma^{\rm H}$ (see Eq.\ (\ref{equ:Sigma_optduenn})).
This possibility is supported
by results which suggest a higher value for the CO-H$_2$ conversion factor
in the outer disks of galaxies and metal-poor environments -- note that the
metallicity in NGC\,4631 decreases with increasing galactocentric radius
(Otte et al.\ \cite{otte+02}). Nevertheless this possibility cannot
explain why $\sigma_{\rm 1.23mm}$ is of the same order or even larger than
$\sigma_{\rm 0.87mm}$,
or the observed excess of the dust emission at $\lambda\,1.23$\,mm.
Another, more speculative scenario, would involve a different way of dust
processing in those areas of NGC\,4631 which are strongly affected by the
interaction.

Therefore a reasonable dust model, whose properties can account for
the mm excess detected above, must also be able to explain the enhanced
values of $\sigma$ at sub-mm/mm wavelengths.

\subsection{What is the origin of the mm excess?}
\label{subsection:verycolddust}

As mentioned above, the sub-mm/mm spectrum of NGC\,4631 is too flat to be
satisfactorily explained with a ``standard'' two-component dust model.
However, several other dust models have been proposed in the past, including
stochastic heating of very small particles, big grains at very low
temperatures ($\ll 10\,{\rm K}$), or grains with different optical
properties, like fractal, fluffy, or ice-coated grains. A good overview
over several models is given by Reach
et al.\ (\cite{reach+95}) in an attempt to interpret COBE observations
of our Galaxy. In the following we summarize the main points of these models
in view of their application to the case of NGC\,4631.

\subsubsection{Very cold big grains}

\begin{figure}
\centering
  \resizebox{\hsize}{!}{\includegraphics[angle=270]{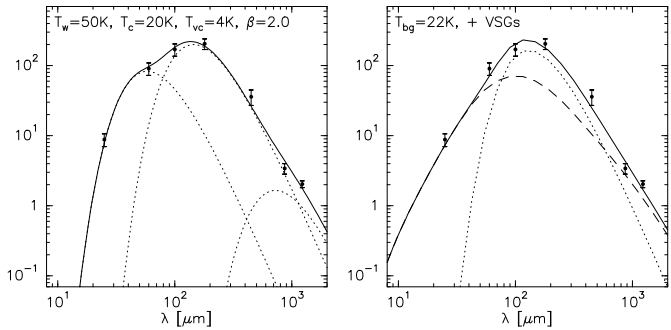}}
  \caption{Two alternative possibilities to fit the FIR-to-mm spectrum of
  NGC\,4631. The left figure shows a possible interpretation including
  a component of very cold dust. The right figure shows a model fit with
  an increased abundance of very small grains. Both possibilities are
  discussed in the text.
  }
  \label{fig:fir}
\end{figure}

In most external galaxies which have been observed in the mm and sub-mm
continuum the cold dust component has a temperature of 15 -- 20\,K.
However, Siebenmorgen et al.\ (\cite{siebenmorgen+99}) have
detected very cold dust in a sample of inactive spiral galaxies with
an average temperature of $\sim 13\,{\rm K}$. These authors also
discuss the possible existence of even colder dust and conclude that
within the optical disk of a galaxy the dust cannot be colder than
about 6\,K.
If we want to explain the mm-excess of NGC\,4631 with such a component of
very cold grains, the most reasonable solution yields a temperature of
$T_{\rm vc} = 4\,{\rm K}$ for this temperature component, and the upper limit
we find is 6\,K.
Such a component has its maximum close to $\lambda \sim 1\,{\rm mm}$ and
could therefore account for the very flat sub-mm/mm part of the flux density
spectrum. The fit to the data with temperatures of 50\,K, 20\,K, and 4\,K
is shown in Fig.\ \ref{fig:fir}.

Such very cold dust could in principle exist in the form of self-shielded
grains in very optically thick clouds with no intrinsic sources.
In order to produce the measured flux density at $\lambda$\,1.23\,mm,
the total dust mass needed is of the order $3 - 6 \times 10^8\,M_{\odot}$
(assuming an absorption coefficient as in the solar neighbourhood),
most of which is at a temperature below 6\,K.
With the gas mass estimated from the CO and \ion{H}{i} data
($8.5 \times 10^9\,M_{\odot}$ including the Helium content) this leads to
a gas-to-dust ratio of $15 - 30$,
far below values in other galaxies or the Milky Way. While this low
ratio could explain the high value we found for the dust absorption cross
section $\sigma$ (see Eq.\ (\ref{equ:sigmakappa})), it is in contradiction
to the low metallicities in NGC\,4631 ($Z/Z_{\odot} \sim 0.5$ on average)
which were found recently by Otte et al.\ (\cite{otte+02}).

On the other hand, Kr\"ugel \& Siebenmorgen (\cite{kruegel+siebenmorgen94})
have shown that in cold dense clouds the dust grains may coagulate to
ice-coated very big grains (up to a radius $a \sim 100\,\mu{\rm m}$), for
which the absorption coefficient can be enhanced by a factor of eight
relative to the diffuse ISM. If a large amount of cold dust existed in this
form, the dust emission could be explained with gas-to-dust ratios
similar to values in the local ISM.

However, in order to keep a large amount of dust at temperatures below
6\,K, compared to the cold component with 15 -- 20\,K, the interstellar
radiation field must be significantly attenuated.
The necessary extinction is of the order $A_V = 15 - 50$ (depending on grain
type), corresponding to a gas column density of $N_{\rm H} \sim 10^{22}$ or
more (see Reach et al.\ \cite{reach+95}, and references therein).
Fig.\ \ref{fig:dist1} shows that values of $N_{\rm H} \sim 10^{22}$
for the total hydrogen column density
are reached only in the inner $3'$ of NGC\,4631, around the maxima of the
CO distribution (assuming a standard CO-H$_2$ conversion factor). While we
would expect radiation-shielded cold dust to exist at places of dense
molecular gas, these regions are also places of strong star formation.
In addition, the mm excess is present over the whole disk of NGC\,4631,
and especially at radii of $1'\!\!.5 - 3'$, where the ratio
$I_{\rm 0.87mm}/I_{\rm 1.23mm}$ is smallest (see Fig.\ \ref{fig:ratsig}).
These areas are located well beyond the molecular ring. 

On the other hand, Galliano et al.\ (\cite{galliano+03}) consistently
explained the mm-excess in NGC\,1569 with very cold grains deeply embedded
in clumps of big grains and primarily heated by the FIR emission of the
latter. This would require a very clumpy medium and small filling factors
of the cold gas and dust.

\subsubsection{Very small grains}

A different model which was successfully applied to the absorption and
emission data from our galaxy over a large wavelength range was presented
by D\'esert et al.\ (\cite{desert+90}). This model included (besides big
grains, which radiate mainly at $\lambda > 70\,\mu{\rm m}$) a component of
amorphous, very small grains (VSGs), which are the main contributor to the IR
emission between 10 and 70\,$\mu$m, and of Polycyclic Aromatic Hydrocarbons
(PAHs), which radiate shortwards of 10\,$\mu$m. The VSGs, which are about
1 -- 20\,nm in size,
are so small that they are heated to non-equilibrium temperatures
by absorption of a single photon and thus show a fluctuating temperature
distribution. They also have a much broader FIR emission spectrum compared
to big grains, and an absorption coefficient proportional to
$\lambda^{-1}$, i.e.\ $\beta = 1$ can be assumed for this component
(Seki \& Yamamoto \cite{seki+yamamoto80}).

Besides our own galaxy, also the FIR-to-mm spectrum of the dwarf galaxy
NGC\,1569 was succesfully fitted with this dust model (Lisenfeld et
al.\ \cite{lisenfeld+02}), although with a stronger ambient radiation
field and a different grain composition (increased abundance of VSGs and
absence of PAHs) than the solar neighbourhood.

We tried to fit the FIR to mm spectrum of NGC\,4631 with the dust model
of D\'esert et al.\ (\cite{desert+90}); the result is shown in
Fig.\ \ref{fig:fir}. For this fit we neglected the PAH component
which radiates at wavelengths smaller than $25\,\mu{\rm m}$.
In order to fit the data, we have to assume an interstellar
radiation field and dust composition different from the solar
neighbourhood.
From the FUV data published by Smith et al.\ (\cite{smith+01}), we estimate
an average ISRF within NGC\,4631 at $\lambda\,1000\,{\rm \AA}$ of about
$6 \times 10^{-2}\,{\rm erg\,cm}^{-2}{\rm s}^{-1}\mu{\rm m}^{-1}$, which
is four times the local ISRF as given in Mezger et al.\ (\cite{mezger+82}).
Compared to big grains, which have an average temperature of $\sim 22\,{\rm K}$
in this model, VSGs are overabundant by a factor of 4 in order to
account for the small $S(0.87\,{\rm mm})/S(1.23\,{\rm mm})$ ratio with
the shallower VSG spectrum in the FIR. In addition, we assume that the VSGs
have sizes up to 20\,nm, compared to 15\,nm as in the Galaxy.
This latter assumption does not necessarily change the
physical properties of the VSGs, like surface-to-volume ratio and heat
capacity (Kamijo et al.\ \cite{kamijo+75}; Stephens \& Russell
\cite{stephens+russell79}).

The application of this model to NGC\,4631 has its drawbacks as well.
The estimated dust mass for the big grain component, assuming standard optical
properties, is $1.1 \times 10^7\,M_{\odot}$, while the VSGs add up to
$0.3 \times 10^7\,M_{\odot}$. The resulting gas-to-dust ratio of
$\sim 600$ is similar to values in extremely metal-poor environments,
and therefore somewhat too high for the metallicities measured in NGC\,4631.
Another critical point is the ISRF of 4 times solar which is required for
a reasonable fit. While this value seems appropriate in the inner part of
NGC\,4631, it is unclear if and how such a radiation field can be maintained
in the outer disk, where the measured absorption cross sections are highest.
Besides these physical concerns, the model fit to the data resulted in a
$\chi^2$ much worse than the model including a component of very cold
grains.

\subsubsection{Grains with unusual optical properties}

Fractal grains have an enhanced efficiency at sub-mm and mm-wavelengths
compared to the optical (e.g. Wright \cite{wright93}). Thus a population
of fractal grains may exist which show rapid temperature fluctuations
(of a few K),
but spend most of the time at very low temperatures due to high
FIR-to-mm emissivity. While the exact properties of this type of grains
are not yet well established, it is known that a much smaller amount of
these grains is needed (compared to compact spherical grains) in order
to produce a comparable amount of emission, hereby explaining the large
values for $\sigma$ found in the previous subsection.
And while the temperature spectrum of these grains may well account for
the measured mm flux density, their mass plays only a minor role in the
total dust mass of
the galaxy. Thus the gas-to-dust ratio could have a reasonable value
and agree with the low metallicity inferred from the Nitrogen
abundance.

If such a component exists, we have to ask why it is detected only in
a few objects. It may be responsible for part of the mm radiation of
the Milky Way (Reach et al.\ \cite{reach+95}) and the dwarf starburst
galaxy NGC\,1569 (Galliano et al.\ \cite{galliano+03};
Lisenfeld et al.\ \cite{lisenfeld+02}), although the
latter authors explained this emission with an overabundance of VSGs.
Many other galaxies do not show any excess in the mm range,
but why should this fractal grain population exist in these
apparently very different objects, but not in many other galaxies of
various types? We can argue that sub-mm continuum observations have
just reached the required sensitivity to detect this effect, and for
many objects only data in the mm {\it or} in the sub-mm regime exist,
while our results show that it is necessary to observe at both
wavelengths to investigate the dust properties of these objects.

After all we should note that in principle strong variations in strength
and colour of the interstellar radiation field may also cause a broadening of
the dust spectrum, hereby potentially flattening the
spectrum in the sub-mm/mm range. However, shortwards of $\lambda$\,0.87\,mm
the spectrum is {\it not} flattened and consistent with a dust absorption
coefficient $\propto \lambda^{-2}$.
Furthermore, the discovered excess in the
mm is quite significant, and with the complexity of the ISM also in
normal galaxies the question arises why the temperature distribution
of dust should be so much narrower in most other galaxies observed up
to date, despite their varying classification and star formation
properties.

\section{Summary and outlook}
\label{section:summary}

We have observed the nearby interacting galaxy NGC\,4631 in the radio
continuum emission at $\lambda\lambda$\,0.87\,mm and 1.23\,mm.
The emission is concentrated on the galactic plane, although some
halo emission could also be detected, especially at 1.23\,mm where
intergalactic cold dust seems to follow the previously detected \ion{H}{i}
spurs. The two maxima visible in both maps, located in the plane
symmetrically around the nucleus at a radius of about $30''$, resemble a
ring-like structure, similar to the distribution of the CO molecular
line emission. In the outer disk, where no CO is detected, the mm/sub-mm
emission follows the \ion{H}{i} distribution, similar to other edge-on
galaxies.

After subtracting non-dust contributions (line, synchrotron, and free-free
emission) from the bolometer flux, the remaining flux densities of
$S_{\rm 0.87mm} = 3.41 \pm 0.58$\,mJy and
$S_{\rm 1.23mm} = 2.04 \pm 0.23$\,mJy
can be attributed to the thermal emission of dust.
In order to estimate the temperature of this dust, we fitted a two-component
modified Planck spectrum to the FIR-to-mm data and found a temperature
of 50\,K for the warm component and of 18\,K for the cold component, which
is responsible for most of the emission at $\lambda > 200\,\mu$m. 

This two-component model suffers from the fact that the observed dust
spectrum is too flat in the sub-mm/mm range. Furthermore it leads to
absorption cross sections too high compared with theoretical expectations
or previous results on other galaxies, especially in the outer part of the
disk which are disturbed because of gravitational interaction.

We suggest several possibilites to resolve these inconsistencies, the most
likely being grains with unusual optical properties which can
account for the measured mm-excess as well as the high absorption cross
sections.

These observations have shown that results of mm and sub-mm observations
have to be combined to investigate the cold dust component in external
galaxies.

\begin{acknowledgements}
We thank the staff of the HHT and the 30-m telescope for their
excellent support, and O. L\"ohmer and N. Neininger for their
collaboration during the observations. Further we would like to thank
E. Kr\"ugel, U. Lisenfeld, C. Popescu, and R. Tuffs for discussions and
comments which helped to improve the paper.
\end{acknowledgements}


\begin{thebibliography}{ }

\bibitem[1998]{alton+98}
Alton, P.B., Bianchi, S., Rand, R.J., et al.\ 1998, ApJ, 507, L125

\bibitem[1999]{alton+99}
Alton, P.B., Davies, J.I., \& Bianchi, S. 1999, A\&A, 343, 51

\bibitem[1974]{andriesse74}
Andriesse, C.D. 1974, A\&A, 37, 257

\bibitem[1999]{baars+99}
Baars, J.W.M., Martin, R.N., Mangum, J.G., McMullin, J.P., \&
Peters, W.L. 1999, PASP, 111, 627

\bibitem[2002]{bendo+02}
Bendo, G.J., Joseph, R.D., Wells, M., et al.\ 2002, AJ, 123, 3067

\bibitem[2003]{bendo+03}
Bendo, G.J., Joseph, R.D., Wells, M., et al.\ 2003, AJ, 125, 2361

\bibitem[1989]{bicay+89}
Bicay, M.D., Helou, G., \& Condon, J.J. 1989, ApJ, 338, L53

\bibitem[1995]{braine+95}
Braine, J., Kr\"ugel, E., Sievers, A., \& Wielebinski, R. 1995, A\&A, 295,
L55

\bibitem[1997]{braine+97}
Braine, J., Gu\'elin, M., Dumke, M., et al.\ 1997, A\&A, 326, 963

\bibitem[1995]{chini+95}
Chini, R., Kr\"ugel, E., Lemke, R., \& Ward-Thompson, D. 1995, A\&A,
295, 317

\bibitem[1991]{rc3}
de Vaucouleurs, G., de Vaucouleurs, A., Corwin, H.G.Jr., et al.\ 1991,
Third reference catalogue of bright galaxies. Springer-Verlag, New York

\bibitem[1990]{desert+90}
D\'esert, F.-X., Boulanger, F., \& Puget, J.L. 1990, A\&A, 237, 215

\bibitem[1990]{devereux+young90}
Devereux, N.A., \& Young, J.S. 1990, ApJ, 359, 42

\bibitem[1984]{draine+lee84}
Draine, B.T., \& Lee, H.M. 1984, ApJ, 285, 89

\bibitem[1997]{dumke+97}
Dumke, M., Braine, J., Krause, M., et al.\ 1997, A\&A, 325, 124

\bibitem[2001]{dumke+01}
Dumke, M., Nieten, Ch., Thuma, G., Wielebinski, R., \& Walsh, W. 2001,
A\&A, 373, 853

\bibitem[1977]{ekers+sancisi77}
Ekers, R.D., \& Sancisi, R. 1977, A\&A, 54, 973

\bibitem[2003]{galliano+03}
Galliano, F., Madden, S.C., Jones, A.P., et al.\ 2003, A\&A, 407, 159

\bibitem[1994]{golla+rw94}
Golla, G., \& Wielebinski, R. 1994, A\&A, 286, 733

\bibitem[1979]{haynes+79}
Haynes, M.P., Giovanelli, R., \& Roberts, M.S. 1979, ApJ, 229, 83

\bibitem[1983]{hildebrand83}
Hildebrand, R.H. 1983, QJRAS, 24, 267

\bibitem[1975]{kamijo+75}
Kamijo, F., Nakada, Y., Iguchi, T., Fujimoto, M.-K., \& Takada, M. 1975,
Icarus, 26, 102

\bibitem[1994]{kruegel+chini94}
Kr\"ugel, E., \& Chini, R. 1994, A\&A, 287, 947

\bibitem[1994]{kruegel+siebenmorgen94}
Kr\"ugel, E., \& Siebenmorgen, R. 1994, A\&A, 288, 929

\bibitem[2000]{lisenfeld+00}
Lisenfeld, U., Isaak, K.G., \& Hills, R. 2000, MNRAS, 312, 433

\bibitem[2002]{lisenfeld+02}
Lisenfeld, U., Israel, F.P., Stil, J.M., \& Sievers, A. 2002, A\&A, 382, 860

\bibitem[2001]{martin+kern01}
Martin, C., \& Kern, B. 2001, ApJ, 555, 258

\bibitem[1982]{mezger+82}
Mezger, P.G., Mathis, J.S., \& Panagia, N. 1982, A\&A, 105, 372

\bibitem[1990]{mezger+90}
Mezger, P.G., Wink, J.E., \& Zylka, R. 1990, A\&A, 228, 95

\bibitem[1999]{neininger+dumke99}
Neininger, N., \& Dumke, M. 1999, Proc.~Natl.~Acad.~Sci.~USA, 96, 5360

\bibitem[1996]{neininger+96}
Neininger, N., Gu\'elin, M., Garc\'{\i}a-Burillo, S., Zylka, R., \&
Wielebinski, R. 1996, A\&A, 310, 725

\bibitem[1995]{niklas+95}
Niklas, S., Klein, U., Braine, J., \& Wielebinski, R. 1995, A\&AS, 114, 21

\bibitem[1997]{niklas+97}
Niklas, S., Klein, U., \& Wielebinski, R. 1997, A\&A, 322, 19

\bibitem[2002]{otte+02}
Otte, B., Gallagher III, J.S., \& Reynolds, R.J. 2002, ApJ, 572, 823

\bibitem[2002]{popescu+02}
Popescu, C.C., Tuffs, R.J., V\"olk, H.J., Pierini, D., \& Madore, B.F. 2002,
ApJ, 567, 221

\bibitem[1994]{rand94}
Rand, R.J. 1994, A\&A, 285, 833

\bibitem[2000]{rand00}
Rand, R.J. 2000, ApJ, 533, 663

\bibitem[1995]{reach+95}
Reach, W.T., Dwek, E., Fixsen, D.J., et al.\ 1995, ApJ, 451, 188

\bibitem[1991]{reuter+91}
Reuter, H.-P., Krause, M., Wielebinski, R., \& Lesch, H. 1991,
A\&A, 248, 12

\bibitem[1988]{rice+88}
Rice, W., Lonsdale, C.J., Soifer, B.T., et al.\ 1988, ApJS, 68, 91

\bibitem[1980]{seki+yamamoto80}
Seki, J., \& Yamamoto, T. 1980, ApSS, 72, 79

\bibitem[1999]{siebenmorgen+99}
Siebenmorgen, R., Kr\"ugel, E., \& Chini, R. 1999, A\&A, 351, 495

\bibitem[2001]{smith+01}
Smith, A.M., Collins, N.R., Waller, W.H., et al.\ 2001, ApJ, 546, 829

\bibitem[1979]{stephens+russell79}
Stephens, J.R., \& Russell, R.W. 1979, ApJ, 228, 780

\bibitem[2000]{stevens+gear00}
Stevens, J.A., \& Gear, W.K. 2000, MNRAS, 312, L5

\bibitem[1988]{strong+88}
Strong, A.W., Bloemen, J.B.G.M, Dame, T.M., et al.\ 1988, A\&A, 207, 1

\bibitem[2000]{thuma+00}
Thuma, G., Neininger, N., Klein, U., Wielebinski, R. 2000, A\&A, 358, 65

\bibitem[1992]{vilacostas+edmunds92}
Vila-Costas, M.B., \& Edmunds, M.G. 1992, MNRAS, 259, 121

\bibitem[1978]{weliachew+78}
Weliachew, L., Sancisi, R., Gu\'elin, M. 1978, A\&A, 65, 37

\bibitem[1988]{werner88}
Werner, W. 1988, A\&A, 201, 1

\bibitem[1993]{wright93}
Wright, E.L: 1993, in: Holt, S.S., Verter F. (eds.), Back to the Galaxy.
AIP, New York

\bibitem[1995]{young+95}
Young, J.S., Xie, S., Tacconi, L., et al.\ 1995, ApJS, 98, 219

\bibitem[1989]{young+89}
Young, J.S., Xie, S., Kenney, J.D.P., \& Rice, W.L. 1989, ApJS, 70, 699

\end{thebibliography}
\end{document}